\documentclass{aa}

\usepackage{graphicx}
\usepackage{txfonts}
\usepackage{subfigure}
\usepackage{float}
\usepackage{natbib,twoopt}
\usepackage{xspace}
\usepackage{balance}
\usepackage{upgreek}
\usepackage{multirow}
\usepackage{mathrsfs}
\usepackage{xcolor}
\usepackage[normalem]{ulem}
\usepackage[colorlinks=true, citecolor=blue]{hyperref}
\usepackage[flushleft]{threeparttable}

\definecolor{redUniBo}{RGB}{187, 46, 41}

\DeclareMathOperator\erf{erf}

\def\apjl{ApJ Lett.}
\def\apjs{ApJ Suppl.}

\def\aap{A\&A}

\newcommand{\Mpch}{$h^{-1}\,\mbox{Mpc}$\,}

\usepackage{placeins}

\bibpunct{(}{)}{;}{a}{}{,} 
\DeclareGraphicsExtensions{.eps,.ps,.pdf}

\newcommand\omegam{0.28}
\newcommand\omegamUp{0.05}
\newcommand\omegamLow{0.04}
\newcommand\seight{0.82}
\newcommand\seightUp{0.14}
\newcommand\seightLow{0.12}
\newcommand\Seight{0.80}
\newcommand\SeightUp{0.08}
\newcommand\SeightLow{0.08}
\newcommand\ascal{0.12}
\newcommand\ascalUp{0.06}
\newcommand\ascalLow{0.06}

\begin{document}

\title{AMICO galaxy clusters in KiDS-DR3: Constraints on cosmological parameters and on the normalisation of the mass-richness relation from clustering}

\author
{G. F. Lesci\inst{\ref{2},\ref{3}}
\and L. Nanni\inst{\ref{1}}
\and F. Marulli\inst{\ref{2},\ref{3},\ref{4}}
\and L. Moscardini\inst{\ref{2},\ref{3},\ref{4}}
\and A. Veropalumbo\inst{\ref{5}} 
\and M. Maturi\inst{\ref{6},\ref{7}}
\and M. Sereno\inst{\ref{3},\ref{4}}
\and \\M. Radovich\inst{\ref{8}}
\and F. Bellagamba\inst{\ref{2},\ref{3}}
\and M. Roncarelli\inst{\ref{2},\ref{3}}
\and S. Bardelli\inst{\ref{3}}
\and G. Castignani\inst{\ref{2},\ref{3}}
\and G. Covone\inst{\ref{9},\ref{10},\ref{11}}
\and \\C. Giocoli\inst{\ref{3},\ref{2},\ref{4}}
\and L. Ingoglia\inst{\ref{9}}
\and E. Puddu\inst{\ref{10}}
}

\offprints{G. F. Lesci \\ \email{giorgio.lesci2@unibo.it}}

\institute{
  Dipartimento di Fisica e Astronomia ``Augusto Righi'' - Alma Mater Studiorum
  Universit\`{a} di Bologna, via Piero Gobetti 93/2, I-40129 Bologna,
  Italy\label{2}
  \and INAF - Osservatorio di Astrofisica e Scienza dello Spazio di
  Bologna, via Piero Gobetti 93/3, I-40129 Bologna, Italy\label{3}
  \and Institute of Cosmology \& Gravitation, University of
  Portsmouth, Dennis Sciama Building, Portsmouth, PO1 3FX, UK\label{1}
  \and INFN - Sezione di Bologna, viale Berti Pichat 6/2, I-40127
  Bologna, Italy\label{4}
  \and Dipartimento di Fisica, Universit\`{a} degli Studi Roma Tre, via
  della Vasca Navale 84, I-00146 Roma, Italy\label{5}
  \and Zentrum f\"ur Astronomie, Universit\"at Heidelberg,
  Philosophenweg 12, D-69120 Heidelberg, Germany\label{6}
  \and ITP, Universit\"at Heidelberg, Philosophenweg 16, D-69120 Heidelberg, Germany\label{7}
  \and INAF - Osservatorio Astronomico di Padova, vicolo
  dell'Osservatorio 5, I-35122 Padova, Italy\label{8}
  \and Dipartimento di Fisica “E. Pancini”, Universit\`{a} di Napoli Federico II, C.U. di Monte Sant’Angelo, via Cintia, I-80126 Napoli, Italy\label{9}
  \and INAF - Osservatorio Astronomico di Capodimonte, Salita Moiariello 16, I-80131, Napoli, Italy\label{10}
  \and INFN - Sezione di Napoli, via Cintia, I-80126, Napoli, Italy\label{11}
}

\date{Received --; accepted --}

\abstract
{
}
{We analysed the clustering of a photometric sample of galaxy clusters selected from the Third Data Release of the Kilo-Degree Survey, focusing on the redshift-space two-point correlation function (2PCF). We compared our measurements to theoretical predictions of the standard $\Lambda$ cold dark matter ($\Lambda$CDM) cosmological model.}
{We measured the 2PCF of the sample in the cluster centric radial range $r\in[5,80]$ $h^{-1}$Mpc, considering 4934 galaxy clusters with richness $\lambda^*\geq15$ in the redshift range $z\in[0.1,0.6]$. A Markov chain Monte Carlo analysis has been performed to constrain the cosmological parameters $\Omega_{\rm m}$, $\sigma_8$, and $S_8 \equiv \sigma_8(\Omega_{\rm m}/0.3)^{0.5}$, assuming Gaussian priors on the mass-richness relation given by the posteriors obtained from a joint analysis of cluster counts and weak lensing. In addition, we constrained the normalisation of the mass-richness relation, $\alpha$, with fixed cosmological parameters.}
{We obtained $\Omega_{\rm m}=\omegam^{+\omegamUp}_{-\omegamLow}$, $\sigma_8=\seight^{+\seightUp}_{-\seightLow}$, and $S_8=\Seight^{+\SeightUp}_{-\SeightLow}$. The constraint on $S_8$ is consistent within 1$\sigma$ with the results from WMAP and Planck. Furthermore, by fixing the cosmological parameters to those provided by Planck, we obtained $\alpha=\ascal^{+\ascalUp}_{-\ascalLow}$, which is fully consistent with the result obtained from the joint analysis of cluster counts and weak lensing performed for this sample.}
{}

\keywords{Galaxy Clusters -- Cosmology: observations -- Cosmology: large-scale structure of Universe -- Surveys}

\authorrunning{G. F. Lesci et al.}

\titlerunning{Cosmological constraints from AMICO KiDS-DR3 cluster clustering}

\maketitle

\section{Introduction}
According to the standard $\Lambda$ cold dark matter ($\Lambda$CDM) cosmological model, galaxy clusters are the largest gravitationally bound systems in the Universe, lying in the highest peaks of the matter density field \citep{article:kaiser1984}. These objects, through their abundances and clustering, trace the statistical properties of the matter density field and its growth \citep{Allen_2011}, providing stringent constraints on cosmological parameters. However, it is difficult to fully exploit the clustering properties of galaxy clusters, due to the complexity of collecting large homogeneous cluster samples. Indeed, surveys of galaxies, probing increasingly larger volumes of the Universe, have played a primary role in this field \citep[see e.g.][]{Guzzo, article:deJong, Aihara}. Ongoing and forthcoming wide extra-galactic surveys, from the lowest to the highest frequencies, will provide complete and pure galaxy cluster samples up to high redshifts and down to low masses. Among them are the Kilo Degree Survey\footnote{\url{http://kids.strw.leidenuniv.nl/}} \citep[KiDS,][]{KIDS,KiDS1000}, the Dark Energy Survey\footnote{\url{ https://www.darkenergysurvey.org}} \citep{des2005,desdr2}, the Vera C. Rubin Observatory LSST\footnote{\url{ https://www.lsst.org/}} \citep{LSST2012,lsst21}, and Euclid\footnote{\url{http://sci.esa.int/euclid/}} \citep{amendola2018, euclDet, baeuclid, scaramella21} in optical and near-infrared, the South Pole Telescope\footnote{\url{https://pole.uchicago.edu/}} \citep{bayliss16,chown18}, the Atacama Cosmology Telescope\footnote{\url{https://act.princeton.edu/}} \citep{act1,act2}, and the Simons Observatory\footnote{\url{ https://simonsobservcatory.org/}} \citep{so1,so2} surveys at high-radio frequencies, as well as eROSITA\footnote{\url{http://www.mpe.mpg.de/eROSITA}} \citep{erosita1,erosita2} in X-rays. \\
\indent Although it is observationally expensive to build up complete and pure samples of galaxy clusters covering wide enough ranges of masses and redshifts, there are numerous advantages in exploiting clusters as cosmic tracers. Galaxy clusters are more clustered than galaxies, with a clustering signal that is progressively stronger for richer systems \citep[see e.g. ][and references therein]{Moscardini_2001}.  Furthermore, clusters are relatively unaffected by nonlinear dynamics at small scales and the redshift-space anisotropies on large comoving scales have a minor impact on the cluster clustering compared to galaxies, thanks to the larger bias \citep[see e.g.][]{article:kaiser,article:hamilton}. Having smaller theoretical uncertainties over the description of nonlinear dynamics and redshift-space distortions, the modelling of the cluster clustering signal is potentially less affected by systematics, compared with the galaxy clustering case. Large galaxy cluster samples have been exploited to provide strong cosmological constraints from both second-order and third-order statistics \citep[see e.g.][and references therein]{Estrada_2009,article:veropalumbo2014, article:veropalumbo2016,article:MarulliXXL,marulli2020, moresco2020,lindholm2020}. These constraints are even more robust when cluster clustering is combined with cluster number counts \citep[e.g.][]{article:mana, article:salvati}.\\
\indent Furthermore, measuring cluster masses gives the opportunity to predict the effective bias of the cluster sample as a function of the cosmological model. It is thus crucial to accurately estimate the cluster masses, possibly with multiple independent probes, such as e.g. X-ray emission or Sunyaev-Zel'dovich effect, and to calibrate mass-observable scaling relations.  In view of the ongoing and next-generation large photometric surveys, which explore the Universe in visible and near-infrared wavelengths, it is decisive to estimate the cluster mass-observable scaling relations involving quantities measured in these bands. Currently, the most reliable mass measurements are provided by weak gravitational lensing \citep[e.g.][]{citlens1,citlens2,citlens3,citlens4,citlens5,citlens6, citesereno, giocoli21, ingoglia22}. Combining this information with the estimate of some optical cluster properties, such as the richness, it is possible to derive cluster mass-observable scaling relations.\\
\indent The goal of this paper is to present  a cosmological analysis based on the monopole of the redshift-space two-point correlation function (2PCF), measured in the catalogue of galaxy clusters developed by \citet{Maturikids}. This catalogue has been built up through the use of the Adaptive Matched Identifier of Clustered Objects (AMICO) algorithm \citep{amico} in the Third Data Release of the Kilo Degree Survey \citep[KiDS-DR3,][]{KIDS}. In this work, we analysed the AMICO KiDS-DR3 catalogue in two redshift bins, assuming a flat $\Lambda$CDM cosmological model, including in the modelling the non-negligible effects of the errors of photometric redshifts (photo-$z$s) on the 2PCF shape. In particular, our reference cosmological model is given by \citet[][Table 2, TT, TE, and EE+lowE]{Planck2018}. Through this analysis we constrained the matter density parameter, $\Omega_{\rm m}$, the amplitude of the density fluctuations, $\sigma_8$, and the {cluster normalisation parameter}, $S_8 \equiv \sigma_8 (\Omega_\text{\rm m}/0.3)^{0.5}$. Furthermore, by fixing the cosmological parameters we inferred the normalisation of the cluster mass-richness scaling relation, finding results consistent with the ones derived from the joint analysis of cluster counts and weak lensing by \citet{lesci2020} in KiDS-DR3. \\
\indent The statistical analyses presented in this paper are performed with the CosmoBolognaLib\footnote{\url{https://gitlab.com/federicomarulli/CosmoBolognaLib/}} (CBL) \citep{cbl}, a set of free software C++/Python numerical libraries for cosmological calculations. Specifically, both the measurements and the statistical Bayesian analyses are performed with the CBL $V5.3$.\\
\indent The paper is organised as follows. In Section \ref{KiDS}, we present the AMICO KiDS-DR3 cluster catalogue. In Section \ref{sec:method}, we describe the methods used to measure and model the 2PCF of this sample. The results of our analysis are presented in Section \ref{2PCF-AMICO}, leading to our conclusions discussed in Section \ref{conclusion}.

\section{Data}
\label{KiDS}
KiDS \citep{dejong13} is an ESO public imaging survey that once completed will cover $1350$ square degrees, exploiting the OmegaCAM wide-field camera \citep{OMEGACAM} on the Very Large Telescope (VLT) survey telescope \citep[VST]{VLT_} with a resolution of $0.21$ arcsec/pixel. This work is based on the catalogue of galaxy clusters detected in KiDS-DR3 \citep{Maturikids} by applying the AMICO algorithm \citep[][discussed in Section \ref{par:AMICO}]{amico}. The KiDS-DR3 catalogue provides the 2 arcsec aperture photometry in the bands $u, \,g,\, r,\, i$, as well as photometric redshifts for all the detected galaxies down to the $5\sigma$ magnitude limits of $24.3,\,25.1,\,24.9$, and $23.8$, for the aforementioned bands, respectively, over a total area of $438$ deg$^2$. The effective area amounts to $377$ deg$^2$, obtained by excluding the regions of the sky affected by satellite tracks and haloes produced by bright stars, or falling in the secondary/tertiary halo masks used for the weak lensing analysis \citep[][]{article:deJong,kuijken15}.

\subsection{The AMICO algorithm}\label{par:AMICO}
AMICO \citep{amico, Maturikids} is an algorithm for the detection of galaxy clusters in photometric surveys, based on the optimal matched filtering technique \citep[see][for a detailed discussion]{article:maturiof}. Specifically, AMICO adopts an iterative approach for the extraction of cluster candidates from the observed galaxy distribution, aiming at maximising the signal-to-noise ratio $(S/N)$ of the overdensity detections, exploiting the statistical properties of both field galaxies and member galaxies, which are described by an arbitrary number of observed quantities. \\
\indent In particular, the detection process adopted in this run of AMICO relies on angular coordinates, $r$-band magnitudes and photo-$z$s of galaxies. The model adopted to describe the clusters is a convolution of a Schechter luminosity function \citep{Schechter} and a Navarro-Frenk-White profile \citep{NFW}. Differently from other cluster detection algorithms based on the so-called red sequence, colours are not considered by AMICO. This has been done to minimize the dependency of the cluster sample on the colour properties of the cluster members. Therefore, AMICO is expected to be effective also at higher redshifts, where the red sequence may be not prominent yet. AMICO has been applied on realistic mock catalogues reproducing the expected characteristics of the future Euclid photometric survey, achieving remarkable purity and completeness levels compared to other algorithms \citep{euclDet}. In fact, AMICO is one of the two algorithms for cluster identification officially adopted by the Euclid mission.
 
\subsection{The AMICO KiDS-DR3 catalogue}\label{sec:catalogue}
\begin{figure}[t!]
\centering
\includegraphics[width=1.\linewidth, height = 6.cm]{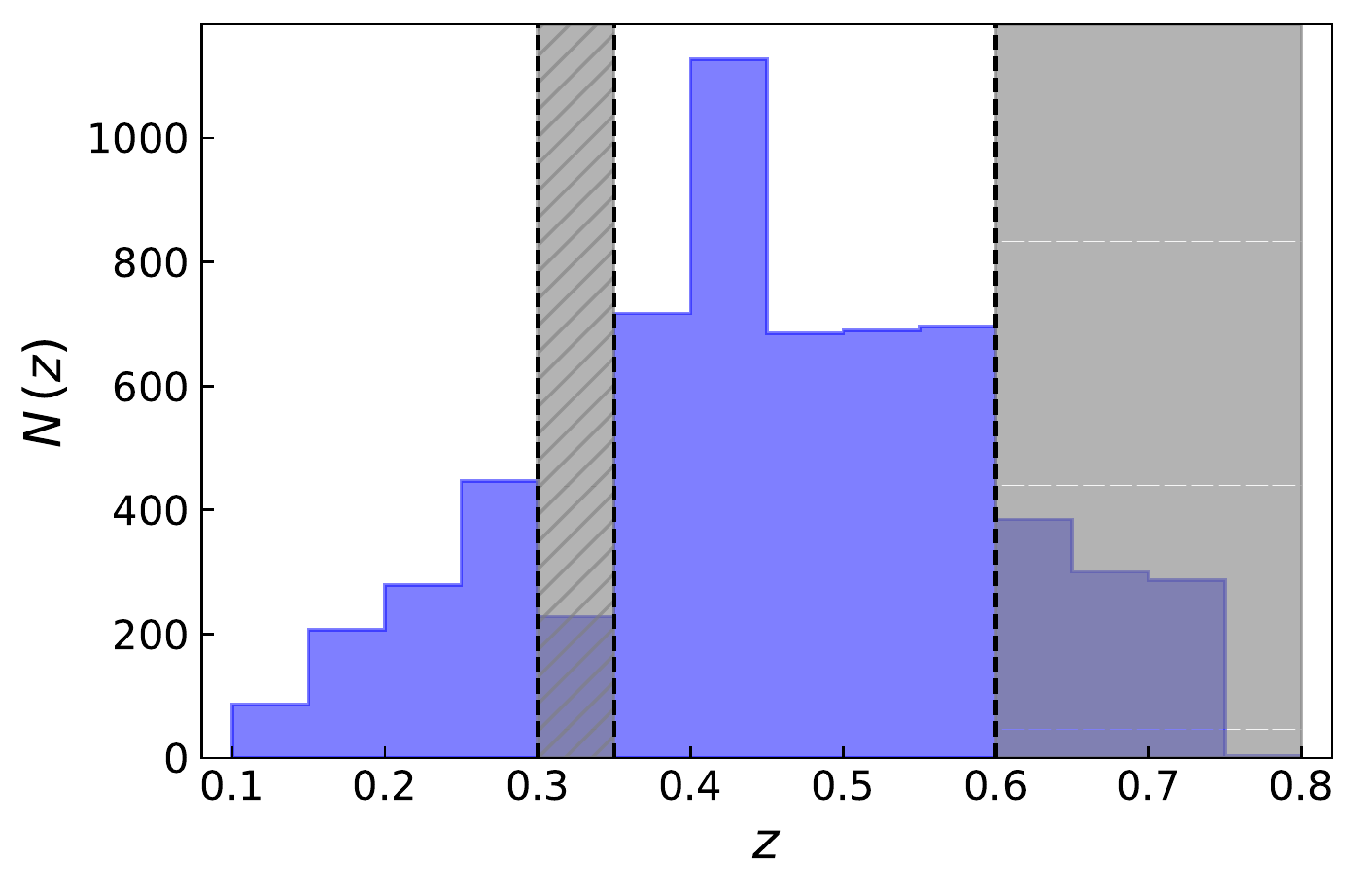}
\includegraphics[width=1.\linewidth, height = 6.cm]{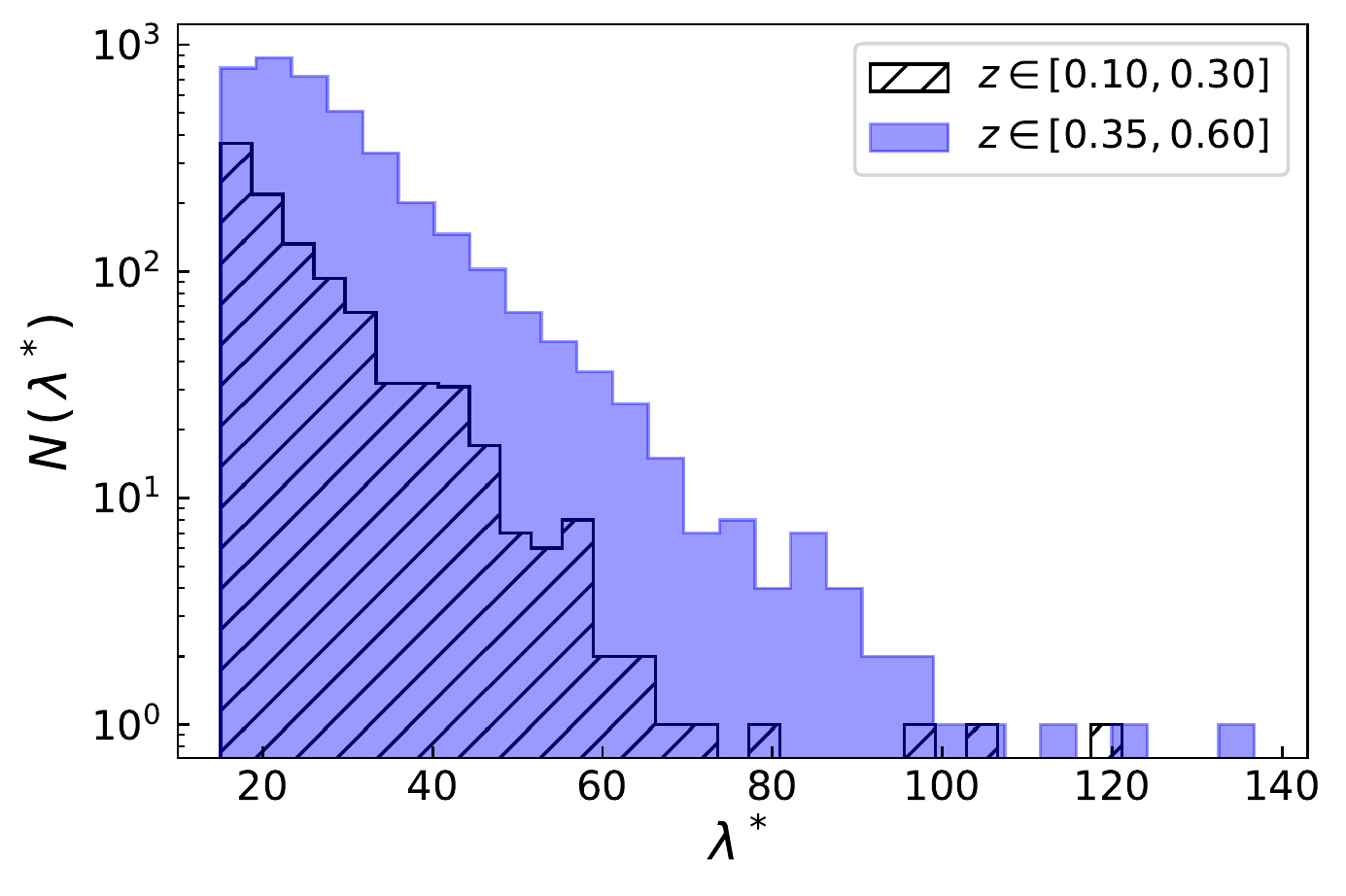}
\caption{Redshift and intrinsic richness distributions of the clusters in the sample. \textit{Top panel:} redshift distribution after the correction for the redshift bias, including only objects with $\lambda^*>15$. The grey shaded areas represent the redshift ranges not used in the analysis. \textit{Bottom panel:} $\lambda^*$ distribution of the objects in the redshift ranges $z\in[0.10,0.30]$ (black hatched histogram) and $z\in[0.35,0.60]$ (blue histogram).}
\label{fig:0}
\end{figure}

We analysed the clustering properties of the AMICO KiDS-DR3 cluster catalogue \citep{Maturikids}, containing 7988 galaxy clusters down to $S/N = 3.5$ and in the redshift range $z\in[0.10,0.80]$. We accounted for the bias described in \citet{Maturikids} affecting cluster redshifts, that is we defined the corrected redshift as $z_{\rm corr} = z - 0.02\,(1+z)$. This bias corresponds to what was found in \citet{KIDS} by comparing the KiDS photo-$z$s to the GAMA spectroscopic redshifts (see their Table 8). Due to the shape of the \textit{g} and \textit{r} filter transmissions, in a small  redshift range around $z\sim 0.32$ the photo-$z$ errors are higher and harder to model. In addition, in the following cosmological analysis (Section \ref{2PCF-AMICO}), we considered the photo-$z$ range $z\in[0.10,0.60]$, since we assumed priors on the mass-richness relation estimated from a weak lensing analysis performed in this redshift range \citep[see][]{article:bellagambalensing}. Consequently, we decided to model the 2PCF in two separate redshift bins: $z\in[0.10,0.30]$,  $z\in[0.35,0.60]$. \\
\indent In the analysis, we used as mass proxy the intrinsic richness, $\lambda^*$, defined as:
\begin{equation}\label{eq:lambda}
\lambda^*_j=\sum\limits_{i=1}^{N_{\rm gal}} P_i(j)\;\;\;\;\text{with}\;\;\;\;
\begin{cases}
m_i<m^*(z_j)+1.5 \\ R_i(j)<{R_{\rm max}(z_j)}
\end{cases}
,
\end{equation}
where $P_i(j)$ is the probability assigned by AMICO to the $i$-th galaxy of being a member of a given detection $j$ \citep[see][]{Maturikids}. The intrinsic richness thus represents the sum of the membership probabilities, that is the weighted number of visible galaxies belonging to a detection, under the conditions given by Eq.\ \eqref{eq:lambda}. The sum of the membership probabilities is an excellent estimator of the true number of member galaxies, as shown in \citet{amico} by running the AMICO algorithm on mock catalogues (see Fig.\ 8 in the reference). In particular, in Eq.\ \eqref{eq:lambda}, $z_j$ is the redshift of the $j$-th detected cluster, $m_i$ is the magnitude of the $i$-th galaxy, and $R_i$ corresponds to the distance of the $i$-th galaxy from the centre of the cluster. The parameter $R_{\rm max}(z_j)$ represents the radius enclosing a mass $M_{200}=10^{14}h^{-1}M_\odot$, such that the corresponding mean density is 200 times the critical density of the Universe at the given redshift $z_j$. Lastly, $m^*$ is the typical magnitude of the Schechter function in the cluster model assumed in the AMICO algorithm. We use the term "intrinsic richness" as opposed to the "apparent richness", both defined in \citet{Maturikids}. In particular, since the threshold in magnitude is always brighter than the survey limit thanks to the redshift dependence of $m^*$, there is no dependence of $\lambda^*$ on the survey limit. Conversely, the apparent richness is a quantity that includes all visible galaxies and therefore is redshift dependent. \\
\indent We considered only the clusters with $\lambda^*>15$, which assures a purity higher than $97$\% over the whole sample \citep[see][]{Maturikids}. Consequently, the sample consists of 1019 clusters for $z\in[0.10,0.30]$ and 3915 clusters for $z\in[0.35,0.60]$. Fig.\ \ref{fig:0} shows the redshift and richness distributions of the objects considered in the analysis.

\section{Methodology}
\label{sec:method}

\begin{figure}
  \centering
    \includegraphics[width=0.95\linewidth, height=14.5cm]{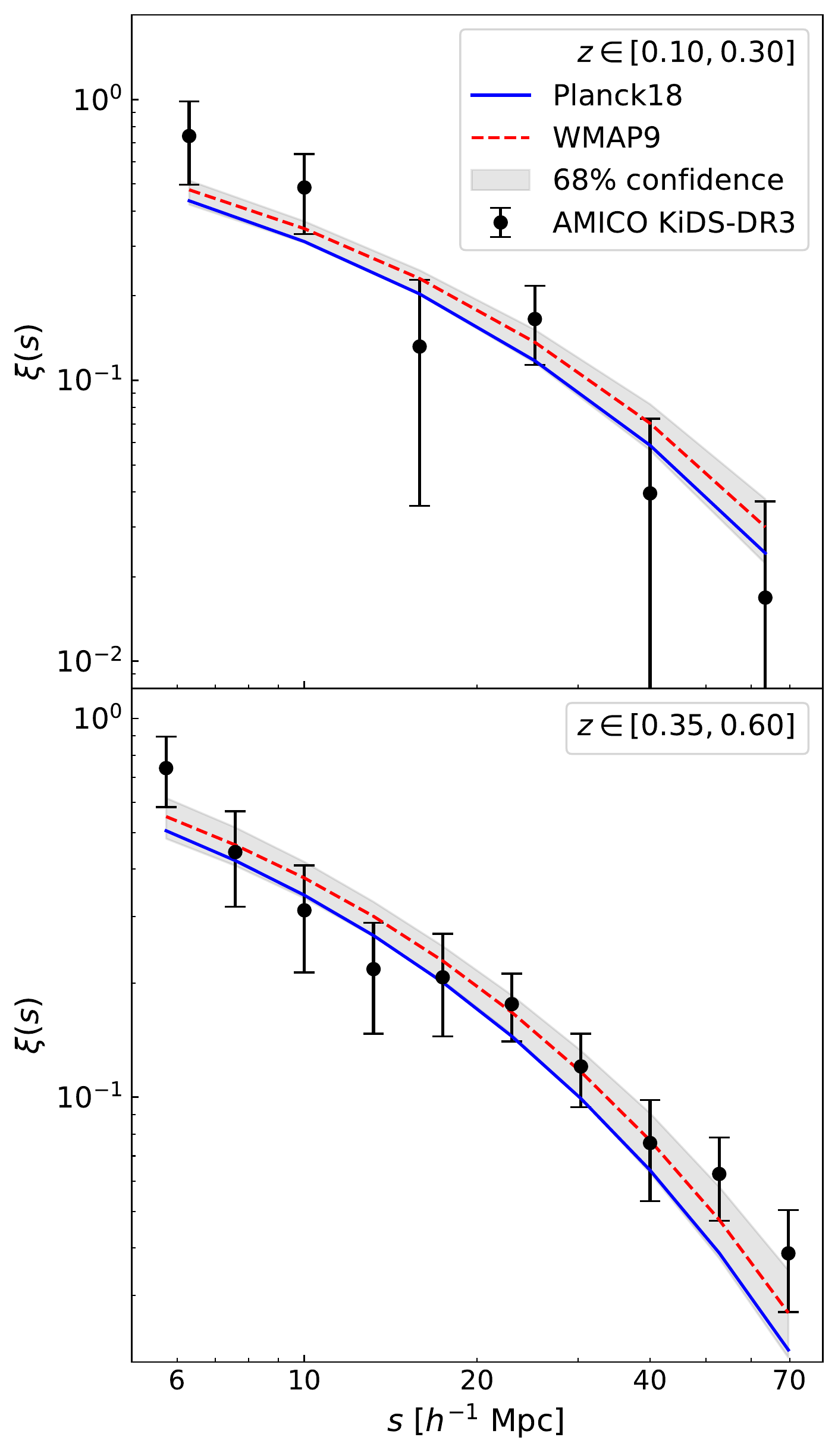}
      \caption{The redshift-space 2PCF (black dots) of the AMICO KiDS-DR3 clusters in the spatial range $s\in[5,80]$ $h^{-1}$Mpc, and redshift ranges $z\in[0.10,0.30]$ (\textit{top panel}) and $z\in[0.35,0.60]$ (\textit{bottom panel}). In both panels, the gray band represents the 68\% confidence level derived from the multivariate posterior of the free parameters considered in the cosmological analysis, described in Section \ref{se:cosmology}. The model computed by assuming the cosmological parameters derived by Planck \citep[][Table 2, TT, TE, and EE+lowE; blue lines]{Planck2018} and WMAP \citep[][Table 3, WMAP-only Nine-year; red lines]{wmap} is represented by the blue solid lines and by the red dashed lines, respectively. In both cases, the median values of the scaling relation parameters derived by \citet{lesci2020} are assumed.}
      \label{fig:2}
\end{figure}

\subsection{Two-point correlation function estimator}
\label{estimator}
We estimate the redshift-space 2PCF with the \citet{article:ls} (LS)
estimator:
\begin{equation}
  \label{eq:2PFCLS}
\xi_{LS}(r) = \frac{N_{RR}}{N_{DD}}\frac{DD(r)}{RR(r)} -
2\frac{N_{RR}}{N_{DR}}\frac{DR(r)}{RR(r)}+1\,,
\end{equation}
where $DD(r)$, $RR(r)$ and $DR(r)$ are the number of data-data, random-random and data-random pairs with separation $r\pm\Delta r$, respectively, while $N_{DD}$, $N_{RR}$ and $N_{DR}$ are the total number of data-data, random-random and data-random pairs, respectively.  This estimator is widely exploited in clustering studies as it is unbiased and with minimum variance for an infinitely large random sample and when $|\xi|\ll1$ \citep{article:hamilton, article:kerscher, article:labatie,keihanen2019}.\\
\indent 
In order to build up the random catalogue, we extracted random (R.A., Dec) cluster positions within the survey tiles by accounting for the same masks used by \citet{Maturikids} for the construction of the cluster catalogue. Regarding the redshifts, we shuffled the observed photo-$z$s. Namely, to each object of the random sample we assigned the photo-$z$ randomly extracted from the AMICO KiDS-DR3 catalogue. Hence, the random sample has the same redshift distribution of the real sample by construction. The generated random catalogue is 30 times larger than the AMICO KiDS-DR3 cluster sample to limit shot noise effects. The observed coordinates are then converted into comoving ones by assuming the $\Lambda$CDM model, with cosmological parameters from Planck18 \citep{Planck2018}.

\subsection{Two-point correlation function model}\label{model}
\begin{table*}[t]
\caption{\label{tab1}Parameters considered in the cosmological analysis.}
  \centering
    \begin{tabular}{l c c r} 
      Parameter & Description & Prior & Posterior \\ 
      \hline
      \rule{0pt}{4ex}
      $\Omega_{\rm m}$ & Total matter density parameter & [0.09, 1] & $\omegam^{+\omegamUp}_{-\omegamLow}$\\ \rule{0pt}{2.5ex}
      $\sigma_8$ & Amplitude of the matter power spectrum & [0.4, 1.5] & $\seight^{+\seightUp}_{-\seightLow}$\\ \rule{0pt}{2.5ex}
      $S_8\equiv\sigma_8(\Omega_m/0.3)^{0.5}$ & Cluster normalisation parameter & --- & $\Seight^{+\SeightUp}_{-\SeightLow}$\\ \rule{0pt}{2.5ex}
      $\alpha$ & Normalisation of the mass-observable scaling relation & $\mathcal{N}(0.04, 0.04)$ & --- \\ \rule{0pt}{2.5ex}
      $\beta$ & Slope of the mass-observable scaling relation & $\mathcal{N}(1.72, 0.08)$ & --- \\ \rule{0pt}{2.5ex}
      $\gamma$ & Redshift evolution of the mass-observable scaling  relation & $\mathcal{N}(-2.37, 0.40)$ & --- \\ \rule{0pt}{2.5ex}
      $\sigma_{\rm intr,0}$ & Normalisation of $\sigma_{\rm intr}$ & $\mathcal{N}(0.18, 0.09)$ & ---\\ \rule{0pt}{2.5ex}
      $\sigma_{\rm intr,\lambda^*}$ & $\lambda^*$ evolution of $\sigma_{\rm intr}$ & $\mathcal{N}(0.11, 0.20)$ & ---\\ \rule{0pt}{2.5ex}
      $\sigma_{z,0}$ & Factor entering the damping of the power spectrum & $\mathcal{N}(0.02, 2\times10^{-4})$ & --- \\ \rule{0pt}{2.5ex}
      $\Omega_{\rm b}$ & Baryon density parameter & $\mathcal{N}(0.0486, 0.0017)$ & ---\\ \rule{0pt}{2.5ex}
      $n_{\rm s}$ & Primordial power spectrum spectral index & $\mathcal{N}(0.9649, 0.0210)$ & --- \\ \rule{0pt}{2.5ex}
      $h\equiv H_0/(100$ km/s/Mpc) & Normalised Hubble constant & $\mathcal{N}(0.7, 0.1)$ & --- \\
    \end{tabular}
  \tablefoot{In the third column, the priors on the parameters are listed. In particular, a range represents a uniform prior, while $\mathcal{N}(\mu,\sigma)$ stands for a Gaussian prior with mean $\mu$ and standard deviation $\sigma$. In the fourth column, we show the median values of the 1D marginalised posteriors, along with the 16th and 84th percentiles. The posterior distributions of $\alpha$, $\beta$, $\gamma$, $\sigma_{\rm intr,0}$, $\sigma_{\rm intr,\lambda^*}$, $\sigma_{z,0}$, $\Omega_b$, $n_s$, and $h$, are not shown since these parameters are not constrained in our analysis.}
\end{table*}
The observed redshift, $z_{\rm ob}$, can be expressed as:
\begin{equation}
z_{\rm ob}=z_{\rm c}+\frac{v_\parallel}{c}\left(1+z_{\rm c}\right)\pm  \sigma_z\,,
\end{equation} 
where $z_{\rm c}$ is the cosmological redshift, $\sigma_z$ is the error on the redshift measurements and $v_\parallel$ is the component of the peculiar velocity along the line-of-sight. Therefore, using $z_{\rm ob}$ to estimate the comoving distance creates distortions in the measures of the 2PCF, not only because of the error on the measurements ($\sigma_z$), but also because $z_{\rm ob}$ encodes information on the peculiar motions along the line-of-sight ($v_\parallel$). The peculiar motions cause the so-called dynamical distortions, an effect commonly known also as redshift-space distortions (RSD).\\
\indent Since our whole analysis is performed on a catalogue extracted from photometric data, taking into account the errors on the observed cluster redshifts is crucial. Following the approach presented in \citet{serenomarulli}, we modelled the redshift-space 2D power spectrum as follows:
\begin{equation}
\label{eq:Pkmu}
P(k,\mu)=P_{\rm DM}(k)\left( b_{\rm eff}+ f \mu^2\right)^2 \, \exp \left( -k^2\,
\mu^2 \, \sigma^2\right)\,,
\end{equation}
where $P_{\rm DM}(k)$ is the dark matter power spectrum, $k=\sqrt{k_\perp^2+k_\parallel^2}$, with $k_\parallel$ and $k_\perp$ being the wave-vector components parallel and perpendicular to the line-of-sight, respectively, $\mu\equiv k_\parallel/k$, $b_{\rm eff}$ represents the effective bias factor (see Section \ref{par:effbias}), $f$ is the growth rate, and the $f\mu^2$ term parameterises the coherent motions of large-scale structure, enhancing the clustering signal at all scales \citep{article:kaiser}. The exponential cut-off term describes the random perturbations affecting redshift measurements, caused by both nonlinear stochastic motions and redshift errors. This is a Gaussian damping term, which causes a scale-dependent effect by removing the signal over a typical scale $k\sim 1/\sigma$, where $\sigma$ represents the
displacement along the line-of-sight due to random perturbations of cosmological redshifts, defined as:
\begin{equation}\label{dampingfactor}
  \sigma \equiv \frac{c\,\sigma_z}{H(z_{\rm m})}\,,
\end{equation}
where $H(z_{\rm m})$ is the Hubble function computed at the mean redshift of the cluster distribution in the bin, $z_{\rm m}$, and $\sigma_z$ is the typical cluster redshift error, expressed as
\begin{equation}\label{eq:sigmaz}
\sigma_z=\sigma_{z,0}(1+z_{\rm m}),
\end{equation}
where $\sigma_{z,0}$ is a free parameter in the analysis. We derived $\sigma_{z,0}$ from the mock catalogue described in \citet{Maturikids}. In particular, we measured the conditional probability $P(z_{\rm ob}|z_{\rm tr})$, where $z_{\rm ob}$ and $z_{\rm tr}$ are the observed and true redshifts, respectively, in several bins of $z_{\rm tr}$, namely $\Delta z_{\rm tr}$. We described the standard deviation of $P(z_{\rm ob}|z_{\rm tr})$ through Eq.\ \eqref{eq:sigmaz}, where $z_{\rm m}$ is the mean value of $z_{\rm tr}$ within $\Delta z_{\rm tr}$. Indeed, given the input galaxy photo-$z$s, AMICO provides unbiased estimates of redshift \citep[see][]{Maturikids}. Then we performed a statistical MCMC analysis assuming a common flat prior on $\sigma_{z,0}$ in all the $\Delta z_{\rm tr}$ bins, obtaining $\sigma_{z,0}=0.02$ with an uncertainty of $\sim2\times10^{-4}$, namely equal to $\sim1\%$. \\
\indent To derive the monopole of the correlation function, we integrate Eq.\ \eqref{eq:Pkmu} over $\mu$, and compute the inverse Fourier transform of the result. The solution can be written as a function of $b_{\rm eff}$ as follows:
\begin{equation}
  \label{eq:miomodello}
  \xi(s)=b_{\rm eff}^2\, {\xi}'(s)+b_{\rm eff} {\xi}''(s)+ {\xi}'''(s)\,,
\end{equation}
where ${\xi}'(s)$ is the inverse Fourier transform of the monopole $P'(k)$, that is:
\begin{equation}
  \label{powerspectrumdamped}
  P'(k)= P_{\rm DM}(k)\frac{\sqrt{\pi}}{2k\sigma}\erf (k\sigma)\,,
\end{equation}
and ${\xi}''(s)$ and ${\xi}'''(s)$ are the inverse Fourier transform of ${P}''(k)$ and ${P}'''(k)$, respectively: 
\begin{equation}
 P''(k)= \frac{f}{\left( k \sigma\right)^3}P_{\rm DM}(k)
  \left[\frac{\sqrt{\pi}}{2}\erf (k\sigma) - k\sigma \exp (-k^2
   \sigma^2) \right] \notag\,,
\end{equation}
\begin{align}
  P'''(k)=& \frac{f^2}{\left( k \sigma\right)^5}P_{\rm DM}(k)
  \frac{3\sqrt{\pi}}{8}\erf (k\sigma) -\\
  &-\frac{k \sigma}{4} \left[ 2
   \left(k\sigma\right)^2+3\right]\exp (-k^2 \sigma^2) \notag\,.
\end{align}

We neglected geometric distortions, which appear when a fiducial cosmology is assumed \citep[in our case,][Table 2, TT, TE, and EE+lowE]{Planck2018} to convert observed coordinates to physical ones, since their effect is negligible with respect to dynamic distortions and photo-$z$ errors \citep[see][]{article:marulli}.

\subsection{Effective bias and mass-richness relation}\label{par:effbias}
The cosmological model of structure formation and evolution predicts that the dark matter halo bias, $b$, primarily depends on halo mass and redshift. Specifically, at a fixed redshift, the bias increases with the tracer's mass, while for a given mass, the bias is an increasing function of the redshift \citep[e.g.][]{article:sheth}. We derived the effective bias in the $i$th bin of redshift, namely $\Delta z_i$, as:
\begin{align}
  \label{eq:biaseff}
  b_{\rm eff}(\Delta z_i) &= \frac{1}{N_i}\sum_{j=1}^{N_i}\int_0^\infty {\rm d}z\,\int_0^\infty{\rm d}\lambda^*\,\int_0^\infty{\rm d}M\,\, b(M,z) \, P(M|\lambda^*,z)\,\,\times \nonumber\\
  &\;\;\;\;\times P(z|z_{{\rm ob},j}) \, P(\lambda^*|\lambda^*_{{\rm ob},j}),
\end{align}
where $N_i$ is the number of clusters in the $i$th redshift bin, $j$ is the cluster index, and $b$ is the halo bias, for which the model by \citet{article:tinker} is assumed. As discussed in Section \ref{se:cosmology}, the results do not significantly change by assuming the halo bias model by \citet{SMT}. In addition, $P(M|\lambda^*,z)$ is a log-normal distribution whose mean is given by the mass-richness scaling relation and the standard deviation (rms) is given by the {intrinsic scatter}, $\sigma_{\rm intr}$, set as a free parameter of the model: 
\begin{equation}
P(\log M|\lambda^*,z)=\frac{1}{\sqrt{2\pi}\sigma_{\rm intr}}\exp\left(-\frac{x^2(M,\lambda^*,z)}{2\sigma^2_{\rm intr}}\right),
\end{equation}
where 
\begin{align}\label{eq:scale_relation}
x(M,\lambda^*,z)=&\log\frac{M}{10^{14}M_\odot/h}\,\,- \nonumber\\
&-\,\Bigg(\alpha+\beta\log\frac{\lambda^*}{\lambda^*_{\rm piv}}+\gamma\log\frac{E(z)}{E(z_{\rm piv})}\Bigg)\,,
\end{align}
where $E(z)\equiv H(z)/H_0$, and we set $\lambda^*_{\rm piv} = 30$ and $z_{\rm piv}=0.35$ following \citet{article:bellagambalensing}. 
In addition, the intrinsic scatter is expressed as
\begin{equation}\label{eq:scatter}
\sigma_{\rm intr} = \sigma_{\rm intr,0}+\sigma_{\rm intr,\lambda^*}\log\frac{\lambda^*}{\lambda^*_{\rm piv}}.
\end{equation}
In Eq.\ \eqref{eq:biaseff}, $P(z|z_{{\rm ob},j})$ and $P(\lambda^*|\lambda^*_{{\rm ob},j})$ are Gaussian distributions, whose mean is the $j$th cluster's observed redshift, $z_{{\rm ob},j}$, and richness, $\lambda^*_{{\rm ob},j}$, respectively. The rms of $P(z|z_{{\rm ob},j})$ is expressed as $\sigma_{z,0}(1+z_{{\rm ob},j})$, where $\sigma_{z,0}$ was derived from the mock catalogue developed by \citet{Maturikids} as described in Section \ref{model}. Analogously, from the mock catalogue we derived an uncertainty on $\lambda^*$ amounting to $\sim17\%$.

\subsection{Likelihood}
\begin{figure*}[t!]
\centering
\includegraphics[width = 0.4 \hsize, height = 7cm] {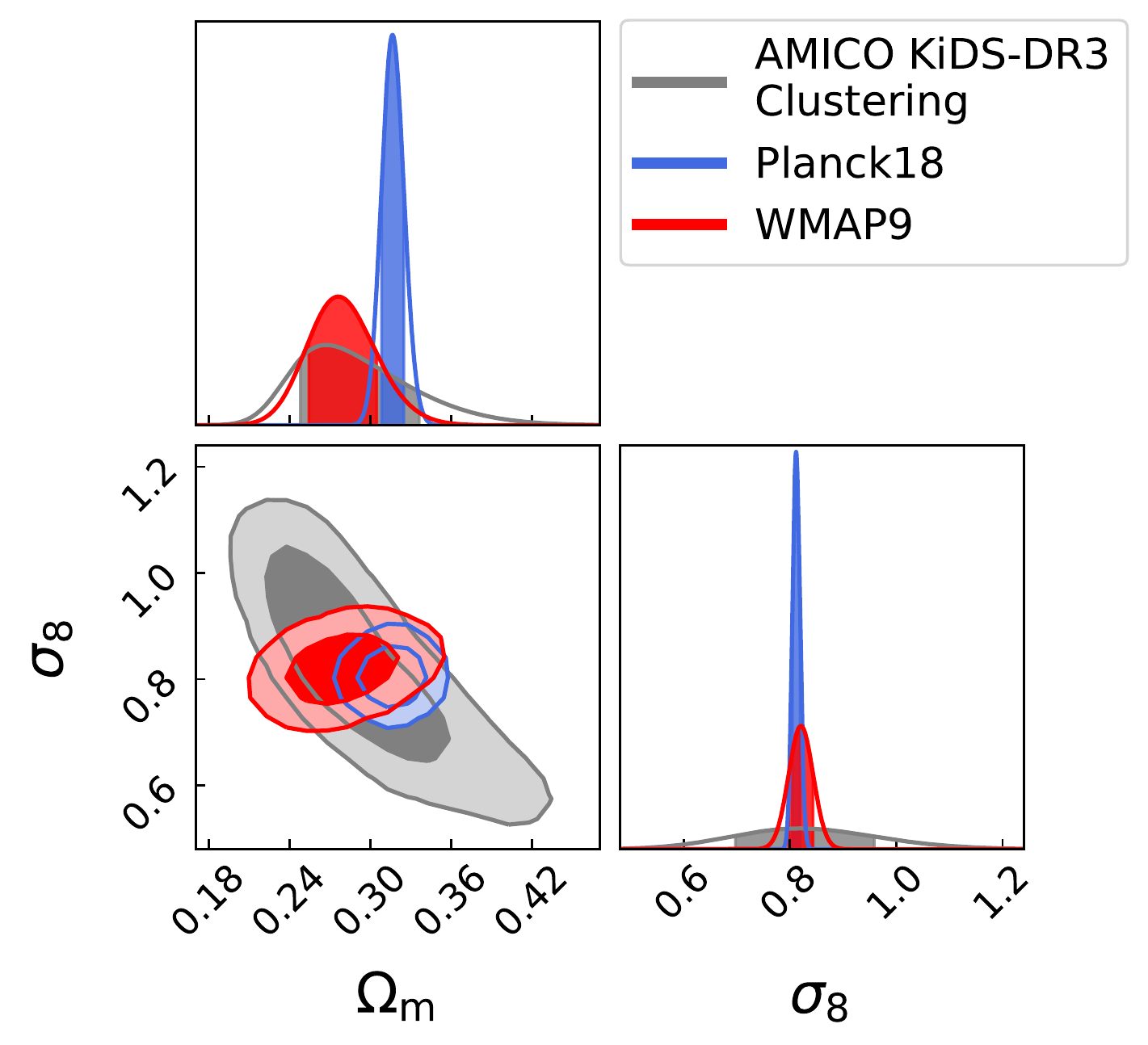}
\includegraphics[width = 0.4 \hsize, height = 7cm] {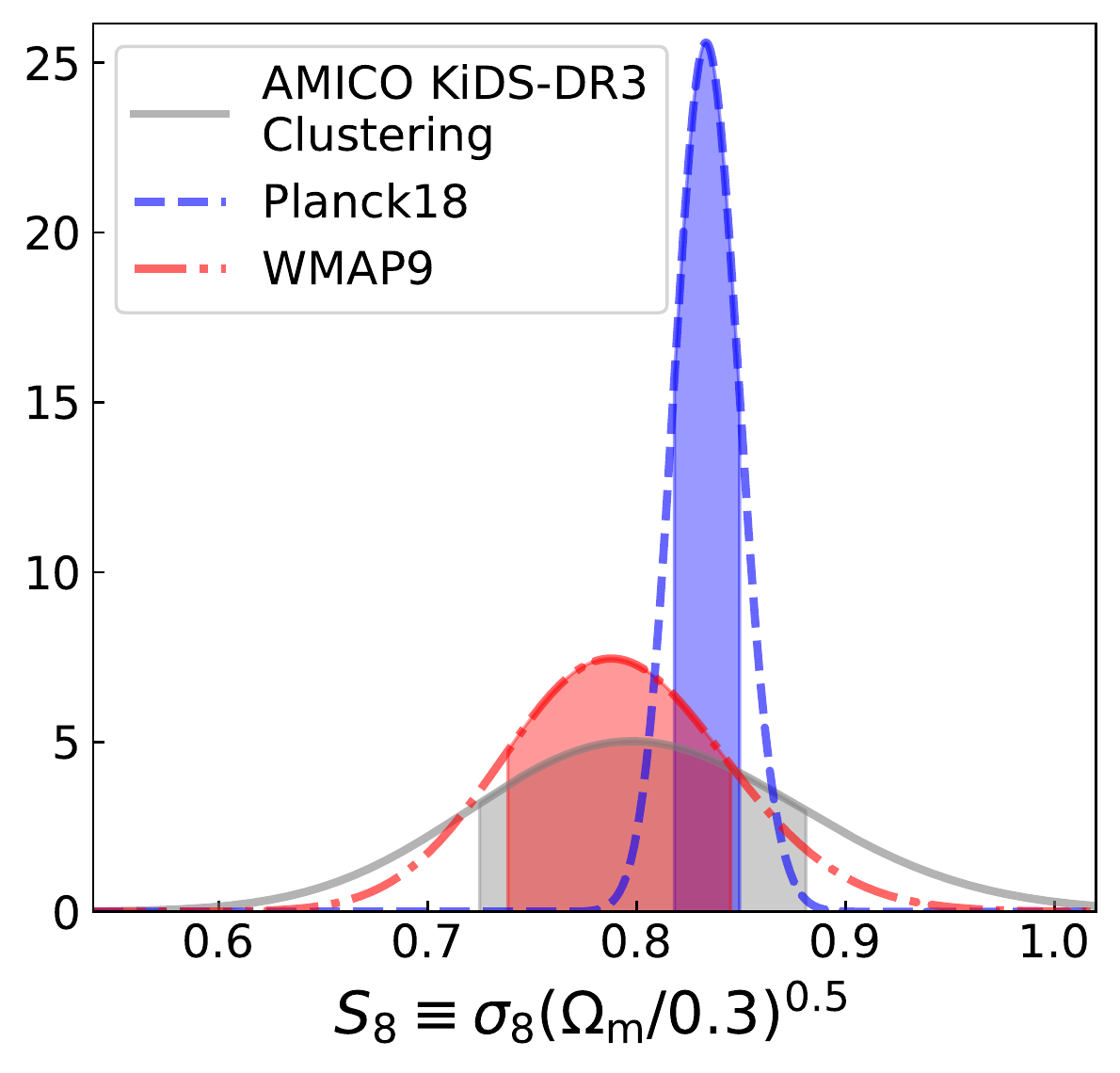}
\caption{Constraints obtained from the cosmological analysis, compared to WMAP and Planck results. In the left panel, we show the 68\% and 95\% confidence levels in the $\Omega_{\rm m}$-$\sigma_8$ parameter space, along with the 1D marginalised posteriors with the relative intervals between the 16th and 84th percentiles, in the case of the cluster clustering analysis of the AMICO KiDS-DR3 catalogue (grey lines). In the same panel, we also display the results from WMAP \citep[][Table 3, WMAP-only Nine-year; red lines]{wmap} and Planck \citep[][Table 2, TT, TE, and EE+lowE; blue lines]{Planck2018}. In the right panel, we show the posteriors for the parameter $S_8$, where the bands show the intervals between 16th and 84th percentiles. The colours are the same as in the left panel.}
\label{fig:results1}
\end{figure*}
For the cosmological Bayesian analysis performed in this work, a standard Gaussian likelihood is considered:
\begin{align}\label{eq:likelihood}
&\mathscr{L} \propto \exp(-\chi^2/2)\,,
\end{align}
with
\begin{align}
&\chi^2=\sum_{i=1}^N\sum_{j=1}^N
  \left(\xi_i^d-\xi_i^m \right)\, C_{i,\,j}^{-1}\, \left(
  \xi_j^d-\xi_j^m\right)\,,
\end{align}
where $N$ is the number of comoving separation bins in which the 2PCF is computed, $d$ and $m$ indicate {data} and {model}, respectively, and $C_{i,\,j}^{-1}$ is the inverse of the {covariance matrix}. The covariance matrix measures the variance and the correlation between the different bins of the 2PCF. It is estimated from the data with the jackknife technique
\citep{article:norberg}:
\begin{equation}\label{eq:covariancematrix}
    C_{i,\,j}= \frac{N_{\rm sub}-1}{N_{\rm sub}}\sum_{k=1}^{N}\left( \xi_i^k -
    \bar{\xi}_i \right)\left( \xi_j^k - \bar{\xi}_j \right),
\end{equation}
where $\xi_i^k$ is the value of the correlation function at the $i$-th bin for the $k$-th subsample, $\bar{\xi}_i$ is the mean value of the subsample, and $N_{\rm sub}$ is the number of resamplings of our cluster catalogue. In particular, the survey tiles are set as the subsample regions for the jackknife.

\section{Results}\label{2PCF-AMICO}
Based on the methods outlined in Section \ref{sec:method}, we performed a cosmological analysis of the redshift-space 2PCF of the AMICO KiDS-DR3 cluster sample. Our analysis is based on two fully independent redshift bins, $z\in[0.10,0.30]$ and $z\in[0.35,0.60]$, and we considered galaxy clusters with $\lambda^*>15$. The 2PCF of such cluster sample is estimated in the spatial range $s\in[5,80]$ $h^{-1}$Mpc. Indeed, at larger scales the clustering signal starts to be weak and dominated by the errors, while at smaller scales the signal is negligible as the cluster size sets the minimum cluster separation. In Section \ref{se:cosmology} we present the clustering measurements and cosmological analysis. The aim of this analysis is to constrain the matter density parameter, $\Omega_{\rm m}$, the square root of the mass variance computed on a scale of 8 \Mpch, $\sigma_8$, and the cluster normalisation parameter, $S_8 \equiv \sigma_8(\Omega_\text{m}/0.3)^{0.5}$, by assuming Gaussian priors for the parameters of the mass-richness relation. In addition, as described in Section \ref{ris:scalingrel}, we investigated a method to infer the normalisation of the cluster mass-observable relation from cluster clustering measurements.

\subsection{Constraints on cosmological parameters}
\label{se:cosmology}
\begin{figure}[t!]
\centering\includegraphics[width = \hsize-0.7cm, height = 12.cm] {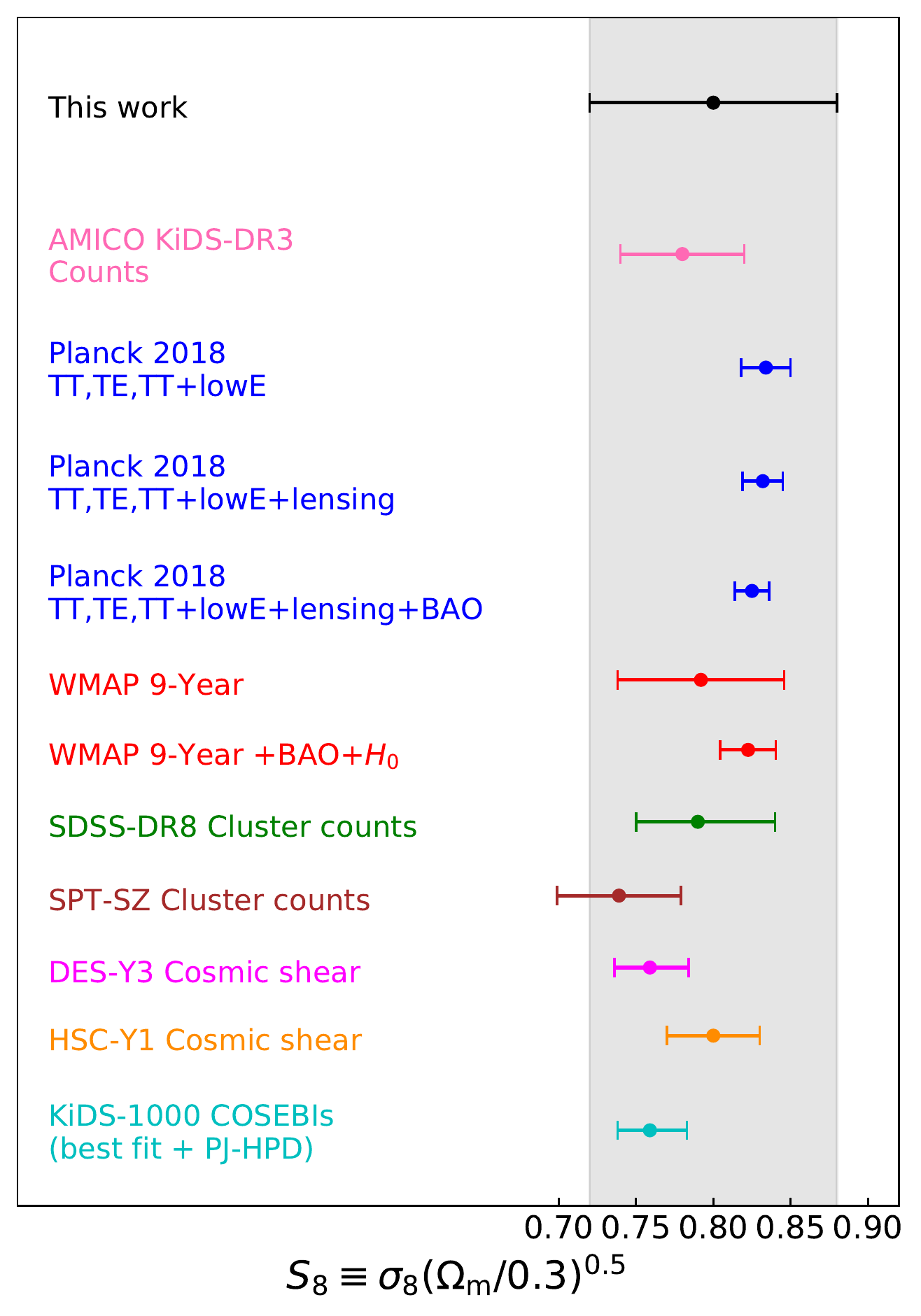}
\caption{Comparison of the constraints on $S_8\equiv\sigma_8(\Omega_\text{m}/0.3)^{0.5}$ obtained, from top to bottom, from the cluster clustering analysis in the AMICO KiDS-DR3 catalogue (black dot), from the joint analysis of cluster counts and weak lensing in KiDS-DR3 performed by \citet{lesci2020} (pink dot), from the results obtained by \citet{Planck2018} (blue dots), \citet{wmap} (red dots), \citet{costanzi} (green dot), \citet{bocquet19} (brown dot), \citet{amon22} and \citet{secco22} (magenta dot), \citet{hsc} (orange dot), and \citet{asgari2020} (cyan dot). The median, as well as the 16th and 84th percentiles are shown.}
\label{fig:results2}
\end{figure}

We exploited the methods described in Section \ref{sec:method} to constrain the main parameters of the $\Lambda$CDM model. We assumed large flat priors for $\sigma_8$ and $\Omega_{\rm m}$, while for the parameters of the mass-richness relation (Eq.\, \ref{eq:scale_relation}, \ref{eq:scatter}), $\alpha$, $\beta$, $\gamma$, $\sigma_{\rm intr,0}$, and $\sigma_{\rm intr,\lambda^*}$, we considered Gaussian priors with mean and standard deviation given by the posteriors derived from the joint analysis of cluster counts and weak lensing performed by \citet{lesci2020}. We also assumed a Gaussian prior on $\sigma_{z,0}$, entering the damping factor of $P(k)$ accounting for the uncertainties on the photo-$z$s (Eq. \ref{dampingfactor} - \ref{eq:sigmaz}), with mean equal to 0.02 and standard deviation equal to $2\times10^{-4}$ (see Section \ref{model}). Lastly, we assumed Gaussian priors for the baryon density, $\Omega_\text{b}$, the primordial spectral index, $n_\text{s}$, and the normalised Hubble constant, $h$, assuming the same mean values derived by \citet[][Table 2, TT, TE, and EE+lowE]{Planck2018}. With regard to the standard deviation of such priors, for $\Omega_\text{b}$ and $n_\text{s}$ we imposed a standard deviation equal to 5 times the 1$\sigma$ error derived by Planck, while for $h$ we assumed a standard deviation equal to 0.1. In Table \ref{tab1} we show the priors and the posteriors of the free parameters of the model. In Fig. \ref{fig:2} we compared our 2PCF measurements in the two selected redshift ranges to the best-fit model. The statistical analysis was performed by assuming a standard Gaussian likelihood, defined in Eq.\ \eqref{eq:likelihood}.\\
\indent We obtained $\Omega_\text{m}=\omegam^{+\omegamUp}_{-\omegamLow}$, $\sigma_8=\seight^{+\seightUp}_{-\seightLow}$, and $S_8=\Seight^{+\SeightUp}_{-\SeightLow}$, where we quote the median, 16th and 84th percentiles, as shown in Fig.\ \ref{fig:results1} and Table \ref{tab1}. Such constraints are in agreement within 1$\sigma$ with WMAP results \citep[][Table 3, WMAP-only Nine-year]{wmap}, and with Planck results \citep[][Table 2, TT, TE, and EE+lowE]{Planck2018}. In addition, Fig.\ \ref{fig:results2} shows that our constraint on $S_8$ is in agreement within 1$\sigma$ with the results obtained from the joint analysis of cluster counts and weak lensing performed in KiDS-DR3 by \citet{lesci2020}. The agreement within 1$\sigma$ holds also for the cluster counts analyses performed by \citet{costanzi}, based on SDSS-DR8 data, and by \citet{bocquet19}, based on the 2500 deg$^2$ SPT-SZ survey data, as well as with the results derived from the cosmic shear analyses performed by \citet{amon22} and \citet{secco22} on DES-Y3 data, \citet{hsc} on HSC-Y1 data, and \citet{asgari2020} on KiDS-DR4 data. In addition, our result on $S_8$ is in agreement with the constraint by \citet{lindholm2020}, namely $S_8=0.85^{+0.10}_{-0.08}$, derived from the autocorrelation of X-ray selected CODEX clusters. We also performed the analysis by assuming the halo bias model by \citet{SMT}, obtaining $S_8=0.79^{+0.08}_{-0.08}$. This result is well in agreement with the one derived from the previously described analysis.

\subsection{The mass-observable scaling relation}\label{ris:scalingrel}
\begin{figure}[t!]
\centering
\includegraphics[width = 0.9 \hsize, height = 7cm] {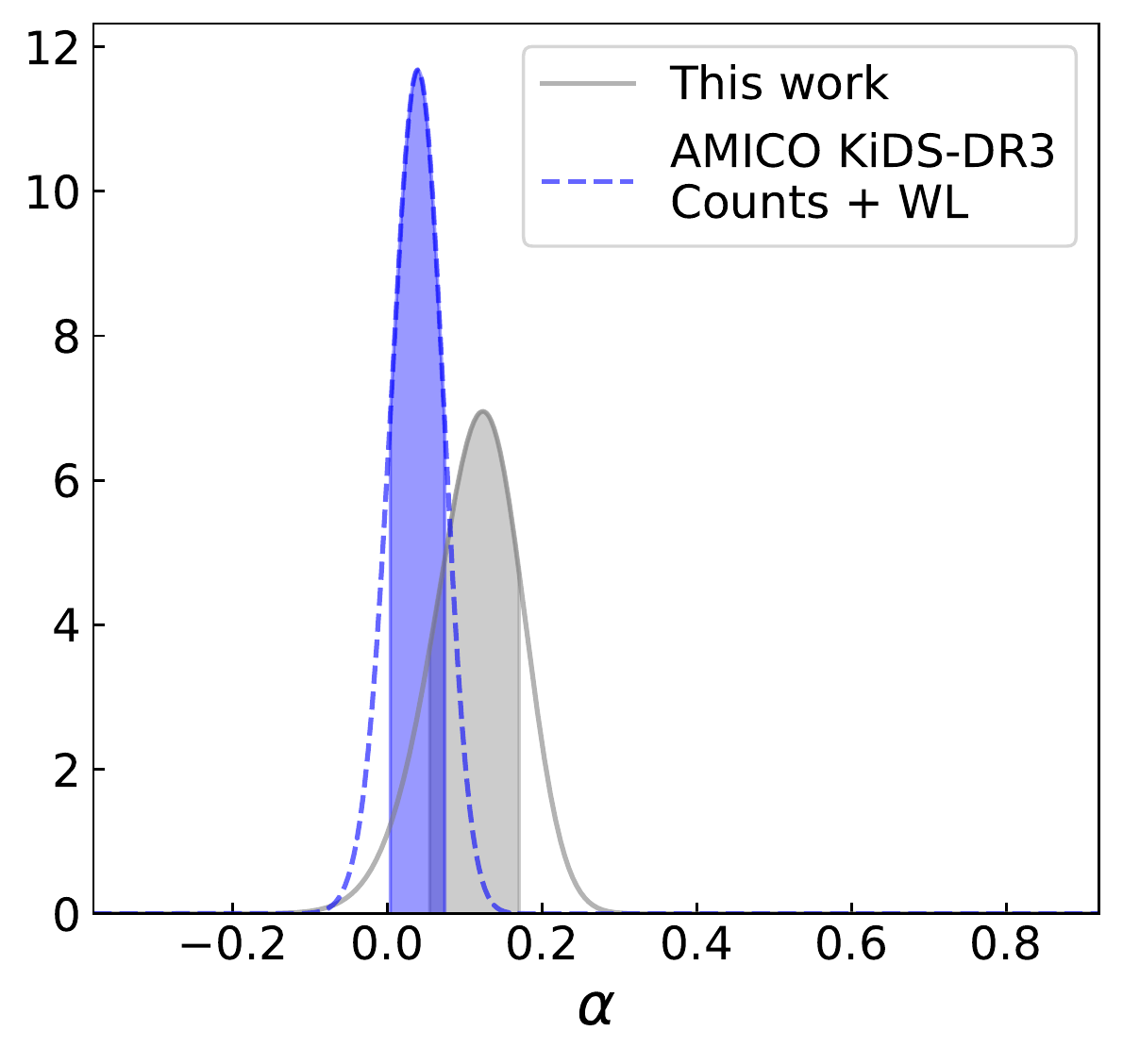}
\caption{Constraint on the normalisation of the mass-richness relation, $\alpha$. The result obtained from the cluster clustering analysis is shown in grey, while the constraint derived from the joint analysis of counts and weak lensing \citep{lesci2020} is shown in blue. The bands show the intervals between 16th and 84th percentiles.}
\label{fig:results3}
\end{figure}

Cluster clustering might provide robust constraints on the normalisation of the mass-observable relation, $\alpha$, if a cosmological model is assumed \citep[see e.g.][]{chiu}. Based on the 2PCF measures used in the cosmological analysis detailed in the previous Section, we also performed the analysis by assuming a flat prior on $\alpha$ and Gaussian priors on $\beta$ $\gamma$, $\sigma_{\rm intr,0}$, and $\sigma_{\rm intr,\lambda^*}$, given by the posteriors derived by \citet{lesci2020}, and the same prior on $\sigma_{z,0}$ assumed in Section \ref{se:cosmology}. In addition, we fixed the cosmological parameters to the values derived by \citet[][Table 2, TT, TE, and EE+lowE]{Planck2018}. We obtained $\alpha=\ascal^{+\ascalUp}_{-\ascalLow}$, which is in agreement within 1$\sigma$ with the result obtained by \citet{lesci2020}, as shown in Fig.\ \ref{fig:results3}. 

\section{Conclusions}\label{conclusion}
In this work, we presented a study of the clustering properties of a photometric sample of galaxy clusters detected by applying the AMICO algorithm on KiDS-DR3 data. The catalogue consists of 4934 clusters in the redshift bins $z\in[0.1,0.3]$, $z\in[0.35,0.6]$, with intrinsic richness $\lambda^*>15$. We measured the monopole of the 2PCF, and performed a cosmological statistical analysis. The clustering model considered includes a damping of the power spectrum to account for the uncertainties on the photo-$z$s. In addition, we derived the effective bias in each redshift bin (Eq. \ref{eq:biaseff}) from the mass-richness scaling relation (Eq. \ref{eq:scale_relation}). \\
\indent We performed a cosmological analysis by assuming flat priors on $\Omega_{\rm m}$ and $\sigma_8$, and Gaussian priors on the parameters of the mass-richness relation given by the constraints derived by \citet{lesci2020}, from the joint analysis of cluster counts and weak lensing in KiDS-DR3. In addition, we marginalised our posteriors over the other cosmological parameters and over the damping of the power spectrum caused by the uncertainties on photo-$z$s. From this modelling, we derived $\Omega_\text{m}=\omegam^{+\omegamUp}_{-\omegamLow}$, $\sigma_8=\seight^{+\seightUp}_{-\seightLow}$, and $S_8=\Seight^{+\SeightUp}_{-\SeightLow}$, which are consistent within 1$\sigma$ with the results obtained from CMB experiments and from state-of-the-art analyses of the evolved Universe. In addition, by fixing the cosmological parameters to the values derived by \citet[][Table 2, TT, TE, and EE+lowE]{Planck2018}, and assuming Gaussian priors on the parameters $\beta$ and $\gamma$ of the mass-richness relation, we derived a robust constraint on the normalisation $\alpha$. In particular, we obtained $\alpha=\ascal^{+\ascalUp}_{-\ascalLow}$, which is in agreement within 1$\sigma$ with the result obtained by \citet{lesci2020}. \\
\indent We expect more accurate constraints on the cosmological parameters and on the mass-richness scaling relation from the analysis of the latest KiDS-DR4 \citep{KiDS1000}, which covers an area of $\sim1000$ square degrees (more than two thirds of the final KiDS area and a factor $\sim2.5$ larger than DR3). KiDS-DR4 has also a photometry which extends to the near infrared (ugriZYJHK$_s$) joining the data from KiDS and VIKING \citep{edge} surveys, which will lead to a better estimate of the photometric redshifts \citep{wright19}: this will be of great help for the application of the AMICO code \citep{AMICOKiDS-1000}, leading to the joint analysis of cluster weak lensing, counts and clustering \citep{LesciDR4}. 

\section*{Acknowledgements}
Based on data products from observations made with ESO Telescopes at the La Silla Paranal Observatory under programme IDs 177.A-3016, 177.A3017 and 177.A-3018, and on data products produced by Target/OmegaCEN, INAF-OACN, INAF-OAPD and the KiDS production team, on behalf of the KiDS consortium.\\
\indent The authors acknowledge the use of computational resources from the parallel computing cluster of the Open Physics Hub (\url{https://site.unibo.it/openphysicshub/en}) at the Physics and Astronomy Department in Bologna. FM and LM acknowledge the grants ASI n.I/023/12/0 and ASI n.2018-23-HH.0. LM acknowledges support from the grant PRIN-MIUR 2017 WSCC32. MS acknowledges financial contribution from contracts ASI-INAF n.2017-14-H.0 and INAF mainstream project 1.05.01.86.10.\\
\indent We thank Shahab Joudaki for his valuable advice.

\section*{Data availability}
The data underlying this article will be shared on reasonable request to the corresponding author.

\bibliography{bib}

\end{document}